\documentclass{article}

\usepackage{authblk,amsmath,amssymb,graphicx,geometry,cite,color,mathptmx,courier,textcomp,verbatim}

\begin{document}

\title{All-Optical Quantum Random Bit Generation from Intrinsically Binary Phase of Parametric Oscillators}

\author{Alireza Marandi$^*$}
\author{Nick C. Leindecker}
\author{Konstantin L. Vodopyanov}
\author{Robert L. Byer}
\affil{E. L. Ginzton Laboratory, Stanford University, CA 94305,}
\affil{marandi@stanford.edu}
\maketitle

{\bf True random number generators (RNGs) are desirable for applications ranging from cryptography to computer simulations. Quantum phenomena prove to be attractive for physical RNGs due to their fundamental randomness and immunity to attack \cite{nature}--\cite{pra11}. Optical parametric down conversion is an essential element in most quantum optical experiments including optical squeezing \cite{squeezing}, and generation of entangled photons \cite{entanglement}. In an optical parametric oscillator (OPO), photons generated through spontaneous down conversion of the pump initiate the oscillation in the absence of other inputs \cite{yariv-siegman, fluorescence}. This quantum process is the dominant effect during the oscillation build-up, leading to selection of one of the two possible phase states above threshold in a degenerate OPO \cite{nabors}. Building on this, we demonstrate a novel all-optical quantum RNG in which the photodetection is not a part of the random process, and no post processing is required for the generated bit sequence. We implement a synchronously pumped twin degenerate OPO, which comprises two identical independent OPOs in a single cavity, and measure the relative phase states of the OPO outputs above threshold as a bit value. We show that the outcome is statistically random with 99\% confidence. With the use of micro- and nanoscale OPO resonators, this technique offers a promise for simple, robust, and high-speed on-chip all-optical quantum random number generators.}

Intrinsic randomness in quantum theory has led to RNG proposals based on the measurement of quantum observables. It was recently demonstrated that quantum correlation of non-local states in two entangled atomic systems can violate Bell inequality principle and be used to generate random numbers \cite{nature}. Entangled-photon  and single-photon sources have been used for RNGs \cite{rayb}. Phase of vacuum fluctuations measured by a homodyne detection scheme \cite{naturephot}, photon arrival time \cite{apl11}, and randomness in photon number distribution of a coherent source \cite{pra11} have also been exploited as the engines of RNGs. Some other quantum RNGs are based on phase measurements of laser diode outputs \cite{jofre, qi} and the Stokes light in Raman scattering \cite{raman}. Most of these methods require massive post-processing to ensure randomness, in some, photodetection is an essential part of the random process, and several require a complicated experimental environment.

The quantum RNG proposed in this letter does not require sophisticated detection techniques because the OPO operates above threshold. Our system is further simplified by using commercially available fiber lasers and well-developed quasi phase matching technology for nonlinear crystals \cite{ppln}.  The optical beam at the output randomly toggles between two intensity levels with a well-defined clock signal. The randomness is intrinsically inherited from the quantum noise of the system, and  we verify minimal contribution of classical noise sources by examining the dependence of randomness on the phase of initiating photons in the OPO. A sequence of 1 billion bits is generated at 10 kbps which has successfully passed all the statistical randomness tests provided by National Institute of Standards and Technology (NIST).

\begin{figure}[htbp]
 \centering
 \includegraphics[width=14cm]{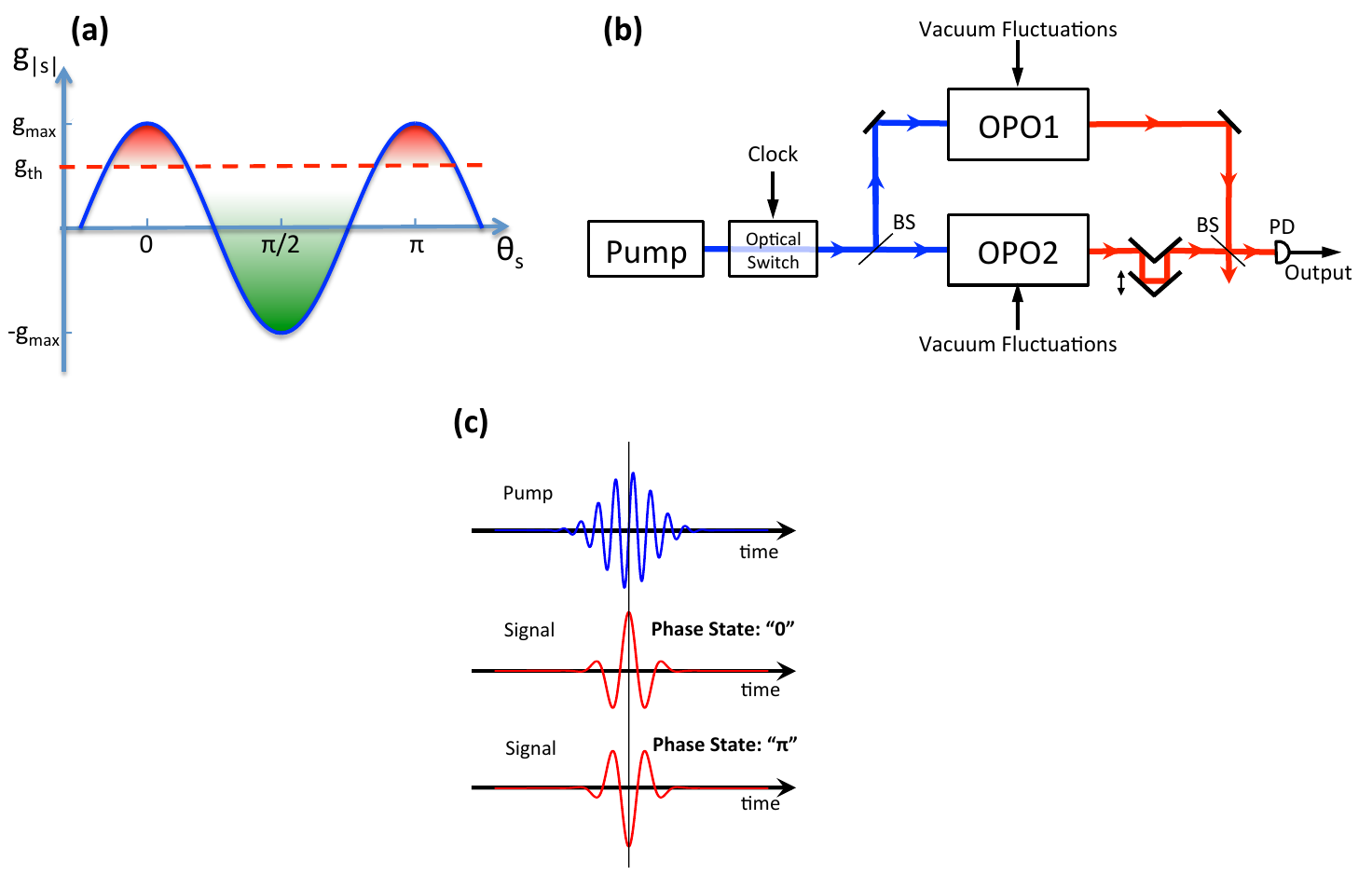}
\caption{a) Phase sensitive gain in a degenerate OPO -- the amplitude incremental gain of the signal ($g_{|s|}$) versus the relative phase between the pump and the signal -- ($\theta_s=\phi_s - \phi_p/2+\pi/4$, where $\phi_s$ and $\phi_p$ are the signal and pump phases, respectively), b) Schematic of the all-optical quantum RNG. PD: Photodetector, BS: non-polarizing beam splitter, c) Electric fields of the signal pulses compared to the pump in the two possible phase states of a short-pulse degenerate OPO.}
\label{fig:sch}
\end{figure}

The operation principle is based on the phase response of a degenerate OPO. In a continuous wave (c.w.) type I degenerate OPO, where signal and idler are indistinguishable (we call it signal in this letter), stable oscillation above threshold can occur in one of the two possible phase states \cite{nabors}. This is because the gain mechanism in such an OPO, provided by degenerate parametric amplification, is phase sensitive and has a period of $\pi$ for the signal phase \cite{yariv}. Figure 1-(a) schematically illustrates the incremental amplitude gain in such an amplifier as a function of the relative phase between the pump and the signal ($\theta_s$). Maximum amplification occurs both at $\theta_s=0$ and $\theta_s=\pi$, where the energy flows from the pump into the signal, and maximum deamplification at $\theta_s=\pi/2$ where the energy flow is from the signal to the pump. Equivalently, this period of $\pi$ exists in the phase dependence of quantum fluctuations in a squeezed state produced by degenerate parametric down conversion \cite{squeezing}. If the maximum gain exceeds the loss in the resonator ($g_{max}>g_{th}=\delta_E$, where $\delta_E$ is the electric field round-trip loss \cite{yariv}), depending on the zero-point fluctuations of the signal modes, the OPO will stably oscillate with the phase of either $\theta_s=0$ or $\theta_s=\pi$. This above-threshold phase state is inherited from the vacuum fluctuations \cite{squeezing}, and none of the design parameters favors oscillation in one or another. 

To make a quantum RNG based on this process, the relative phase states between two identical OPOs can be measured by interfering the outputs. This is demonstrated in the schematic in Fig. 1-(b), where two degenerate OPOs are pumped with the same source, and an optical switch is used to restart the OPOs periodically. If both path lengths are matched, the interference of the output signals will randomly toggle between high and low intensity levels resulting in a sequence of ``zeros" and ``ones" at the photodetector output.

\begin{figure}[htbp]
 \centering
 \includegraphics[width=14cm]{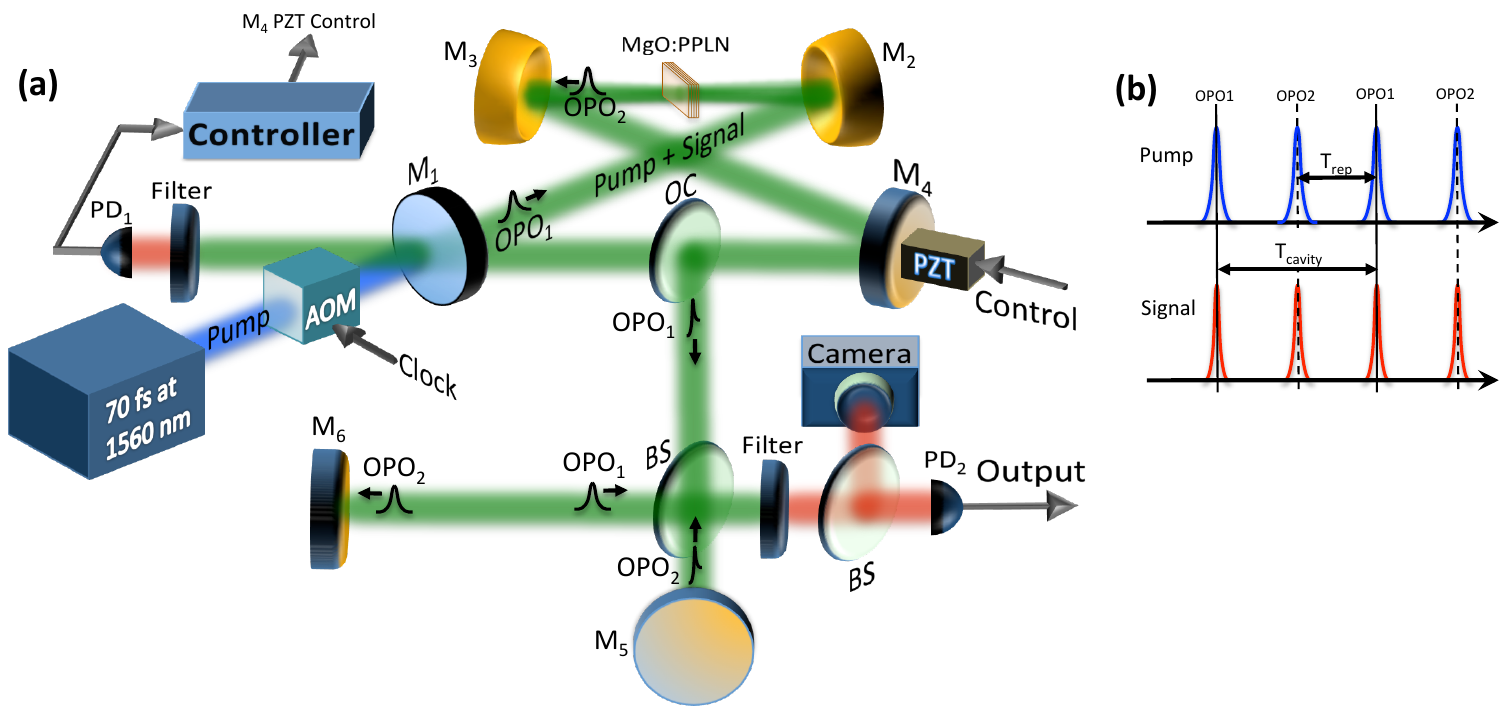}
\caption{ a) Schematic of the experimental setup consisting of the twin degenerate OPO followed by an un-equal arm Michelson interferometer. The round-trip length difference of the arms in the interferometer is equal to the separation length of pump pulses. The output of the interferometer is filtered to eliminate the pump. An AOM is used to restart the twin OPO at a clock rate of 10 kbps, while the zeroth order output (non-diffracted beam) pumps the OPO. The detector PD1 and a piezoelectric transducer (PZT) on M$_4$ are parts of the ``dither-and-lock" servo loop for stabilizing the cavity length to a resonance \cite{cleo11}. The interferometer arm length difference is stabalized using the interference of the undepleted pump and a PZT on M$_6$ which is not shown in this schematic. OC: Output Coupler, b) Pump and signal pulse trains for a twin OPO in which the cavity roundtrip time ($T_{cavity}$) is twice the repetition period of the pump ($T_{rep}$); this results in two independent OPOs with signal pulse trains at half the repetition frequency of the pump, marked by OPO1 and OPO2 on the pulse trains. }
\label{fig:sch}
\end{figure}

We realized the OPO-based quantum RNG using the synchronously pumped femtosecond degenerate OPO demonstrated in \cite{opex}. For such a short-pulse OPO, similar to the c.w. case, the signal can take one of the two possible phase states \cite{sam} which destructively interfere with each other. Fig. 1-(c) depicts an example of the electric field of the pump and the signal in the two possible phase states. The existence of these phase states for a degenerate femtosecond OPO was experimentally verified in \cite{cleo11}.

To avoid implementation of two matched resonators, a twin OPO is used, in which two identical OPOs operate in the same ring resonator with the roundtrip path length twice the separation length of pump pulses, as depicted in Fig. 2-(a). Fig. 2-(b) shows the pump and signal pulse trains for this twin OPO. The pump pulses generate two independent signal pulse trains, marked as OPO1 and OPO2 in the figure. These two temporally separated OPOs have half the repetition rate of the pump, the same polarization and spectral properties, and experience exactly the same optical paths in the oscillator. An unequal arm interferometer is used to measure the relative phase states of OPO1 and OPO2 by interfering their temporally separated pulse trains, as depicted in the schematic of Fig. 2- (a).

\begin{figure}[htbp]
 \centering
 \includegraphics[width=12cm]{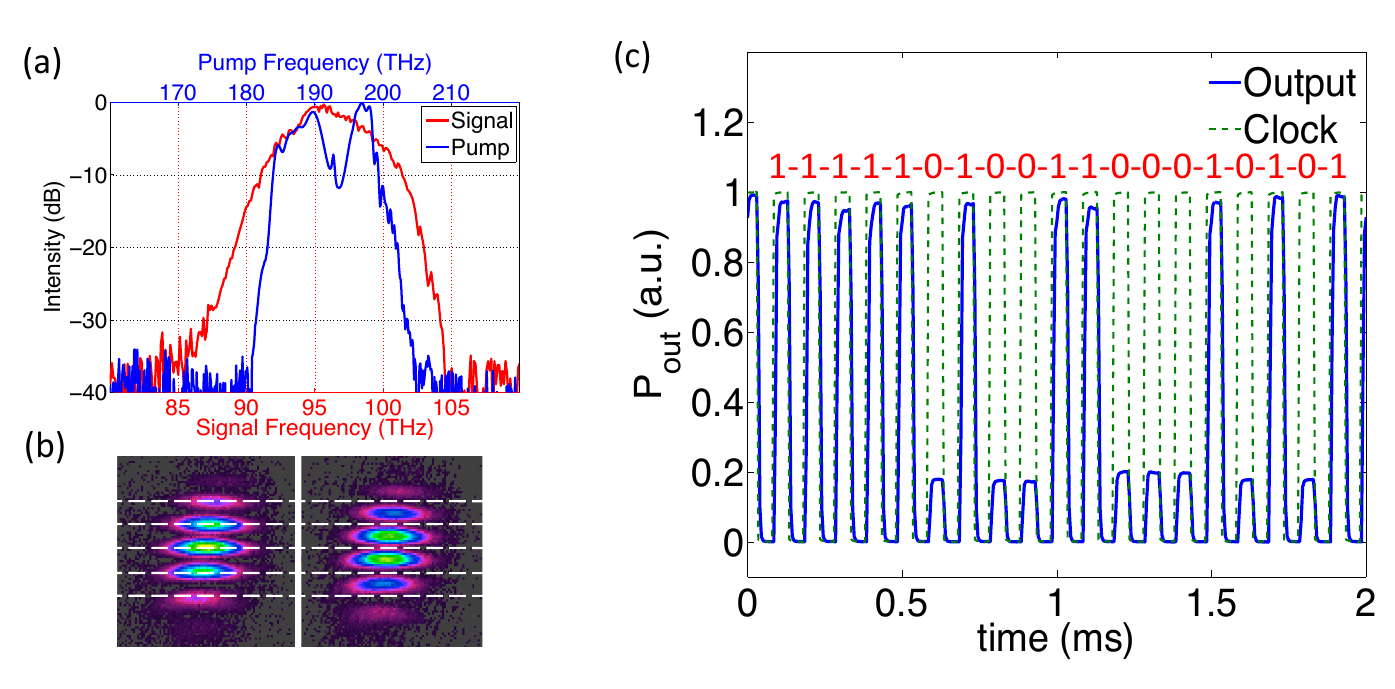}
\caption{a) Optical spectrum of the pump and signal, b) complementary fringe patterns at the output of the Michelson interferometer, i.e. maximum on one pattern is minimum on the other and vice versa, c) sample time domain signal of the interferometer output, i.e. the random sequence, along with the clock signal at 10 kbps.}
\end{figure}

The OPO stabilized with the servo loop could run continuously for several days (see Methods). Figure 3-(a) shows the signal spectrum centered at 3.1 $\mu$m and the pump centered at 1.56 $\mu$m. Complementary stable fringe patterns at the output of the interferometer were obtained as depicted in Fig. 3-(b) when the beams in the arms are slightly angled vertically. Blocking and unblocking the pump resulted in random toggling between these two patterns.  

To capture a bit stream, the beam angles are well-aligned in the interferometer and a photodetector is used at the output, while an Acousto-Optic Modulator (AOM) causes periodic restarting of the twin OPO. Figure 3-(c) depicts a sample of the signal at the output of the interferometer along with the clock signal at 10 Kbps applied to the AOM. The binary sequence is extracted from the interferometer output as depicted in the plot.

\begin{figure}[htbp]
 \centering
 \includegraphics[width=8cm]{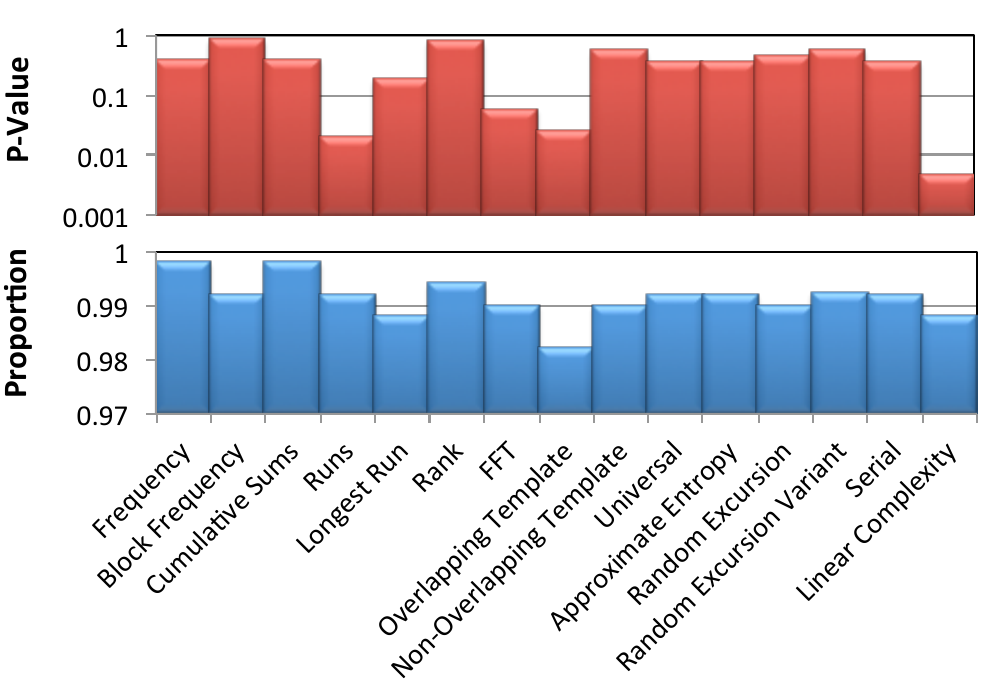}
\caption{Summary of the results of the NIST statistical tests \cite{nist} on a sequence of 1 billion bits. The sequence is chopped into 500 shorter sequences and each test is performed on all sequences. Horizontal axes shows the name of the tests, $P-Value$ is the result of $\chi^2$ distribution evaluation for all short sequences, and $Proportion$ is the proportion of the short sequences passed each test with a significance level of 0.01. The final decision on if the whole sequence has passed a test is made based on these criteria: $P-Value >0.0001$ and $Proportion > 0.976$ \cite{nist}, which are satisfied for all the tests. }
\end{figure}

A sequence of 1 billion bits is taken with this method, and it proves to be random with an average of 0.5000. To verify the statistical randomness, a series of tests developed by NIST \cite{nist} are performed, and the summary of results are presented in Fig. 4. The 1-Gb sequence passed all the NIST statistical tests indicating it is random with 99\% confidence.

\begin{figure}[htbp]
 \centering
 \includegraphics[width=14cm]{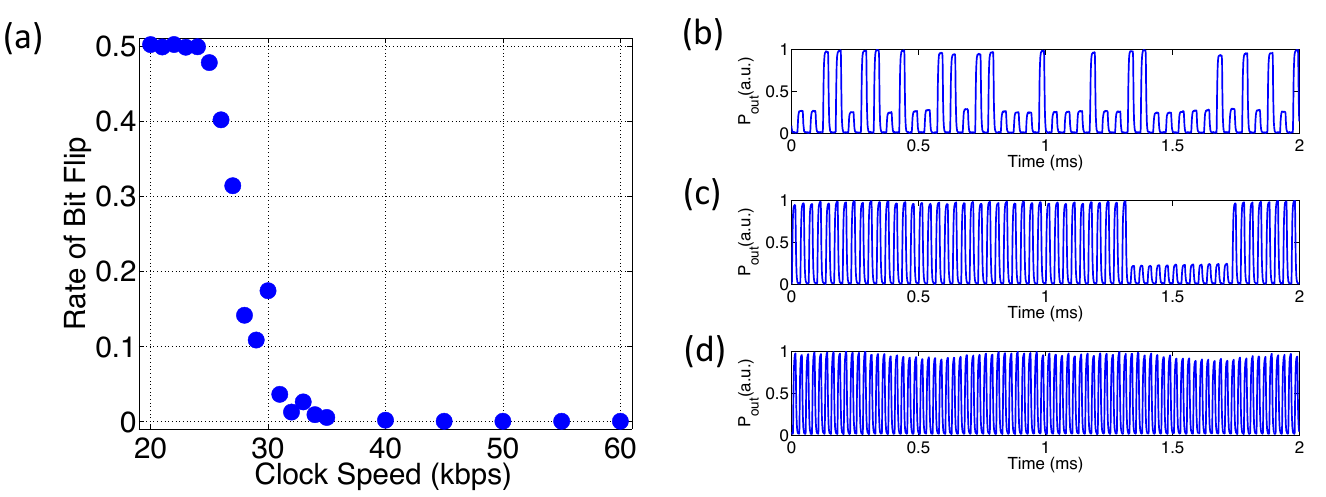}
\caption{ a) Dependence of the randomness on the clock speed; for each clock speed a sequence of 100 kb is recorded, and the rate of bit flip is calculated by counting the number of changes in the bit value and dividing it to the sequence length,  for a random sequence the rate of bit flip is close to 0.5 and for a single-value sequence it is zero; The transition from random to non-random output occurred around 25 kbps which is close to the estimated value of 30 kbps taking into account the experiment parameters. This value can be increased by decreasing the overall pump power and approaching the threshold. To record the 1 Gb of data, the pump power was reduced so that the randomness would not break until 200 kbps (the measurement limit). b, c, d) a sample of the output sequence when clock speed is b) 20 kbps, c) 30 kbps, and d) 40 kbps. }
\end{figure}

The maximum bit-rate supported by this RNG depends on the turn-``on" and turn-``off" dynamics of the OPO. At the end of each clock cycle, the intracavity field must decay to the quantum noise level \cite{yariv-siegman}, or the residual field from the previous state will seed oscillation of the next state, and randomness will be lost. In the design presented here, the clock speed is slow enough (and OPO ``on"-time long enough) that oscillation build to a steady-state level.  In this case, the clock rate is limited by the time it takes for the intensity to decay below noise, which is 10 to 20 times longer than the 1/e cavity decay time when the OPO is pumped well over threshold and allowed to reach steady state. Therefore, the RNG speed is limited by the cavity decay time, and faster bit rates, in the Gbps range are expected to be achievable using pumps with higher repetition rates, and shorter OPO cavities \cite{81ghz,pplnWG}.

However, one can operate the OPO closer to threshold, where the build-up time is longer than the decay time. In this case, the oscillation will not necessarily reach steady-state, but relative phase may still be measured with a sufficiently sensitive detector to establish the bit value, decoupling maximum RNG speed from the cavity decay time. A potential extreme case is eliminating the cavity and having a single pass parametric down conversion with a speed as high as the repetition rate but requiring either a sensitive detection system or a relatively high peak power. It is even possible to design a similar RNG with a c.w. pump to eliminate the repetition rate limit.

We exploit the OPO dynamics discussed above to verify that the randomness is due to the phase of photons initiating the oscillation. As mentioned before, our RNG operates in the regime where build-up is much faster than decay. This is due to the low ($\sim$23\%) modulation depth of the AOM that biases the pump only slightly below threshold during the ``off" phase of the clock, resulting in a much longer decay time than if the OPO were un-pumped. This makes it easy to observe the transition to a fixed-state, `latched' oscillation, when at faster clock rates the turn ``on" edge arrives while many photons remain in the cavity. Figure 5-(a) shows how the randomness -- measured here by the rate of bit flip in a 100-kb long sequence -- breaks as a function of clock speed. Figure 5-(b),(c),(d) illustrate output samples for three points on this curve. This experiment shows that when an intracavity phase reference exists -- in the form of residual photons from the previous clock cycle -- the randomness of the sequence breaks and the influence of classical noise sources is negligible.

 In summary, we demonstrated a novel technique for all-optical quantum RNGs based on the randomness in spontaneous parametric down conversion by measuring the above-threshold phase of oscillation in a degenerate OPO. The twin system, made of two identical OPOs sharing the same cavity, facilitates the phase state measurement. Stable operation is achieved using a free running commercially available mode-locked fiber laser. Unlike many other quantum RNGs, this system enables quantum random number generation with no need for electronic or computer post processing on the generated bit sequence, and has the potential of extremely fast operation. Implementation of an OPO-based quantum RNGs is not limited to second-order nonlinear materials and free space elements as presented here; one can extend this approach to on-chip $\chi^3$ OPOs \cite{gaeta}--\cite{vlasov} where CMOS compatibility can open new possibilities for new generation of RNGs for future communication and computational systems. Combined with micro- and nano resonators, this technique paves the way for high-speed all-optical quantum RNGs with multi Gigabit per second rates.

\vspace{5 mm}
{\bf Acknowledgements}

The authors would like to thank S. E. Harris, Y. Yamamoto, M. M. Fejer, C. R. Phillips, J. S. Pelc, C. W. Rudy, and O. Crisafulli for invaluable discussions, and K. Urbanek for experimental support.

\section{Methods}
\subsection{Experimental Setup}

The OPO is pumped by a 1560-nm mode-locked Er-fiber laser (Menlo Systems C-fiber, 100 MHz, 70 fs, 300 mW)  where beam is conditioned by a mode-matching telescope for efficient pumping of the OPO.  The resonator is a 6-m ring cavity corresponding to twice the spacing between the pump laser pulses.  The cavity optics comprise one pair of concave mirrors with ROC=50 mm (M$_2$ and M$_3$), and six flat mirrors (twice folded bow tie), five of which are gold coated with approximately 99\% reflection  (only one is depicted in schematic of Fig. 2: M$_4$ ).  A single dielectric mirror  (M$_1$) is used to introduce the pump, which has 90\% transmission for the pump and more than 99\% reflection in the 2.8 - 4 $\mu$m. This mirror has a `chirped' design of dielectric layers to compensate the dispersion of the nonlinear crystal. 

Broadband  gain centered around 3.1 $\mu$m is provided by 1-mm long MgO-doped periodically poled lithium niobate (MgO:PPLN) crystal. The poling period is 34.8 $\mu$m for broadband  type-0 (e=e+e) phase matching at a temperature of 32$^\circ$C. The crystal is cut such that the mid-IR beam propagates perpendicular to the poling domains when the beam enters at the Brewster angle. The beam waist for signal beam in the crystal is \texttildelow10 $\mu$m. The mirrors M$_2$ and M$_3$ are set to 5-degree angle of incidence to compensate the astigmatism caused by the Brewster angled crystal and allow stable resonances in the 6-m long cavity. The output is extracted with a pellicle beam splitter (OC) having \texttildelow 8\% reflection over a broad bandwidth. The filters are AR coated Ge substrates to block the pump and allow the mid-IR signal go through.

Oscillation occurs when signal/idler waves are brought into degenerate resonance by fine-tuning the cavity length with the piezo stage of M$_4$.  Three of these resonances occur separated by \texttildelow1.5 $\mu$m of roundtrip cavity length, corresponding to half of the signal central wavelength \cite{opex}. Continuous operation of the OPO is obtained by locking the cavity length to track the center of the strongest resonance using a dither-and-lock scheme \cite{opex}. The twin OPO starts oscillating at a pump average power of about 120 mW, and the maximum mid-IR output power is 4 mW. 

\subsection{Estimated Maximum Speed}
The twin OPO operates in the regime where the build up time is faster than the decay time. Therefore, for quantum RNG outcome the turn off time should be long enough to allow the intracavity power to decay from the steady state level, that is about 1 $W$, to the quantum noise level, i.e. one photon per mode ($P_{noise}=h\nu \Delta \nu$ \cite{yariv}), which is about 1 $\mu W$. Here $h\nu$ is the photon energy at the central signal wavelength of 3.1 $\mu$m, and $\Delta \nu$ is the OPO bandwidth at 3-dB level, estimated to be $\sim$10 THz (Fig. 3-a ). The intensity decay time of the OPO can be estimated using: $$ \tau_{off}=\frac{T}{2\delta_E-2\delta_E \sqrt{\frac{P_{off}}{P_{th}}}},$$ where $\delta_E$ is the electric-field fractional round-trip loss, $P_{th}$ is the pump power at threshold, $P_{off}$ is the pump power at the ``off" state, and $T$ is the cavity roundtrip time. In the presence of the AOM, the OPO threshold measured before M$_1$ is increased to 190 mW because of pulse broadening in the AOM. The pump power at the off state is 168 mW, and intracavity power loss ($2\delta_E$) is estimated to be 0.27 resulting in the 1/e intensity decay time of 1.2 $\mu$s. Hence the minimum turn-off time required for decaying from steady state power to quantum noise level is about 17 $\mu$s corresponding to a maximum clock speed of $\sim$30 kbps.

\end{document}